\documentclass[journal]{IEEEtran}
\usepackage{graphicx}
\usepackage{subfigure}
\usepackage{indentfirst}
\usepackage{amsmath}
\usepackage{caption2}
\usepackage{cite}
\usepackage{algorithm}
\usepackage{algorithmic}
\usepackage{amsthm}
\renewcommand{\IEEEiedlistdecl}{\settowidth{\labelwidth}{W}}
\begin{document}
\title{A scientific understanding of network designing}
\author{Guoqiang Zhang
            \thanks{Guoqiang Zhang is with the Institute of Computing Technology, Chinese Academy of Sciences, Beijing, China. Email: guoqiang@ict.ac.cn}
            }
\markboth{IEEE/ACM Transactions on Networking}{Guoqiang Zhang: A
Scientific Understanding of Network Designing}
 \maketitle

 \begin{abstract}
As the Internet becomes severely overburdened with exponentially
growing traffic demand, it becomes a general belief that a new
generation data network is in urgent need today. However, standing
at this crossroad, we find that we are in a situation that lacks a
theory of network designing. This issue becomes even more serious as
the recent progress of network measurement and modeling challenges
the foundation of network research in the past decades.

This paper tries to set up a scientific foundation for network
designing  by formalizing it as a multi-objective optimization
process and quantifying the way different designing choices
independently and collectively influence these objectives. A
cartesian coordinate system is introduced to map the effect of each
designing scheme to a coordinate. We investigated the achievable
area of the network designing space and proved some boundary
conditions. It is shown that different kind of networks display
different shapes of achievable areas in the cartesian coordinate and
exhibit different abilities to achieve cost-effective and scalable
designing. In particular, we found that the philosophy underlying
current empirical network designing and engineering fails to meet
the cost-effective and evolvable requirements of network designing.
We demonstrated that the efficient routing combined with effective
betweenness based link bandwidth allocation scheme is a
cost-effective and scalable design for BA-like scale-free networks,
whereas if other designing choices cannot be determined beforehand,
ER network is a markedly good candidate for cost-effective and
scalable design.

 \end{abstract}


 \section{Introduction}
\IEEEPARstart{T}{he} Internet becomes more and more complex and
susceptible to congestion today. With the emergence of new
data-intensive applications and fast growing population in need of
data communication service, most experts agree that the existing
data network architecture is severely stressed and approaching its
capability limits\cite{routing-scaling, huston-1, huston-2,
claffy2003}. Thus the move to a brand new next generation data
network is in urgent need today.

However, before this move, there lacks the theory of network
designing, which becomes more serious as the recent surge of network
measurement and modeling efforts show quite counter-intuitive
result, that is, most networks, including the Internet, display
scale-free structures rather than presumably random
structures\cite{powerlaw1999,BA,GLP,dk-series}. This recent trend
poses an urgent need as well as an opportunity for a fundamental
understanding of the science underlying the network designing. The
Network Science and Engineering Council(NetSE) has recommended the
creation of ``a fundamental science of networking that has the
potential to underpin systematic engineering methodologies for the
design of \emph{large-scale}, \emph{robust}, \emph{cost-effective}
and \emph{evolvable} networked systems'' to be among the top agenda
for long-term network research \cite{netse}. The hardness of
establishing this science primarily lies in the abstraction of
network designing. Too simple model may lose its practicality, while
too complicated model that includes too many details may lead to
insolvability. At the heart of various designing considerations, a
scientific approach should clarify the following key issues: what
are the typical network designing objectives and major designing
choices, what are the inherent interplays of these choices, how the
designing choices independently and collectively affect different
designing objectives, and what are the achievable values of the
designing objectives.

From the macroscopic perspective, network designing often involves
several independent yet closely related aspects, of which the three
most fundamental ones are what kind of network topology shall be
used, which routing algorithm is effective, and how the link
bandwidth will be allocated. One can independently choose either
ingredients as they are conceptually orthogonal to each other.
However, when network designing objectives are concerned, these
ingredients become closely interleaved because different
combinations will produce different effects.

One major objective for network designing and engineering is to
enhance the network's transmission capacity, a focus of a number of
previous researches\cite{gupte-1, gupte-2,
edge-deletion,efficient-routing,onset-traffic-congestion,node-capability-redistribution
}. However, we argue that most often, maximizing the network
transmission capacity could not be the sole objective. A network
designing is also typically constrained by the financial or
technical restrictions, so a \emph{cost-effective} design is
preferable. In addition, as networks are constantly evolving, a
designing scheme with good \emph{scalability} will be more
desirable. To summarize, previous researches have the following
shortcomings : (1) they focus only on one aspect of the network
designing ingredients, without a formally defined scientific
framework ; (2) their studies are based on a traffic flow model that
is too simple for network engineering; (3) they don't consider the
technical feasibility of the designing scheme, which is crucial for
network engineering; (4) they do not consider the scalability issue
of a designing scheme, which is of significance in reality because
networks are constantly evolving.

Motivated by these incentives, in this paper, we carried out an
in-depth study of network designing in a systematic approach. The
main contributions are:
\begin{enumerate}
\item We propose that network designing is a multi-objective
optimization process, with often contradictive objectives, and
formalize the framework for network designing study by defining a
two dimensional cartesian coordinate system according to the two
important designing objectives;
\item We compare different designing schemes by quantitatively mapping the effect of each
scheme to a coordinate in the coordinate system, and we explore the
achievable areas in this coordinate system;
\item We investigate the cost-effectiveness and scalability of some
representative designing schemes. In particular, we find that the
current network designing philosophy is neither cost-effective nor
scalable. We propose one cost-effective and scalable designing
scheme, i.e., efficient routing combined with effective betweenness
based bandwidth allocation, which can achieve pretty good tradeoff
between the two designing objectives for BA-like scale free
networks, and we demonstrate that if one has the full freedom to
select network designing choices, ER network is a remarkably good
candidate to achieve a cost-effective and scalable design,
especially when other designing ingredients cannot be determined
beforehand.

\end{enumerate}

The subsequent paper is structured as follows: We briefly review
related work in Section {\ref{related-work} and introduce our
modified traffic flow model in section \ref{model}. The major
network designing objectives and designing choices are presented in
section \ref{designing-objective-choice}. We move forward to the
definition and analytical analysis of the cartesian coordinate
system and achievable areas in section \ref{coordinate-system}. The
scalability issue of network designing is discussed in section
\ref{scalability}. Finally, we conclude the paper in section
\ref{conclusion}

\section{Related Work}\label{related-work}
The past decade has witnessed a surge of network topology
measurement and modeling research activities\cite{powerlaw1999,BA,
clustering, assortative-mixing, ark, DIMES,
rocketfuel,chinese-internet, internet-evolution}, which, with
abundant evidence, show that most real networks, including the
Internet, display scale-free network structure. This milestone
finding challenges the foundation of network related studies, since
for decades the network researches are grounded on the general
assumption of random network model \cite{erdos59}.

Recently, there is a research shift from the static network topology
characterization and modeling to the traffic dynamics on these new
networks\cite{polymorphic-torus,gupte-1, gupte-2,
edge-deletion,efficient-routing,onset-traffic-congestion,node-capability-redistribution}.
These work, primarily for enhancing the network's transmission
capacity, can be broadly classified into three categories: to make
small changes to the underlying network
topology\cite{gupte-1,gupte-2,edge-deletion}, to adjust the routing
algorithm\cite{efficient-routing} and to customize the node
capability\cite{onset-traffic-congestion,
node-capability-redistribution}. A straightforward way to improve
the network's transmission capacity is to create new edges in the
network\cite{gupte-1, gupte-2}. A somewhat counter-intuitive
approach proposed in \cite{edge-deletion} is to enhance the BA
network's transmission capacity by removing edges that connect nodes
with high betweenness values, hence balancing the traffic across the
network. In \cite{efficient-routing}, the authors try to achieve the
same goal by the so called efficient routing algorithm. Instead of
routing packets along the path with minimum number of hops,
efficient routing routes packets along the path that minimizes the
sum of node degrees, hence bypassing those high degree nodes in BA
network. In \cite{onset-traffic-congestion}, the authors investigate
this issue from yet another angle, i.e., to adjust the node
capability. Two node capability models are studied in this work,
i.e., with node capability proportional to its degree and
betweenness respectively. These two capability models are further
investigated under the condition that the total capability remains
fixed\cite{node-capability-redistribution,
cost-effective-designing}, which essentially turns the problem of
node capability assignment to an optimization problem. A central
concept behind these analysis is the betweenness
centrality\cite{freeman-betweenness,
centrality-and-network-flow,load-distribution,optimal-network-topology},
which bridges the research of static network topology and dynamic
network traffic analysis. Under certain circumstances, the
betweenness centrality precisely estimates the traffic passing
through a node or an
edge\cite{centrality-and-network-flow,load-distribution,optimal-network-topology,efficient-routing}.

\section{Traffic flow model}\label{model}
 In this study, we use a traffic-flow model that is slightly
 different from
\cite{efficient-routing, edge-deletion,
onset-traffic-congestion,Danila06} to more accurately model the
network realism. Each node is capable of generating, forwarding and
receiving packets. At each time step, $R$ packets are generated at
randomly selected sources. The destinations are also chosen
randomly. Instead of assigning a single node capability to each node
as previous studies do, we assign multiple link bandwidths to
multiple interfaces of each node. Denote $C(v_i)$ to be the link
bandwidth of the $i$th interface of node $v$, then $C(v_i)$
specifies the maximal number of packets this interface can deliver
at a time step. We assume symmetric link bandwidth, i.e., for an
edge $e=(v_i, u_j)$, we have $C(v_i)=C(u_j)$, which is further
denoted as $C(e)$.  Each packet is forwarded toward its destination
based on the particular \emph{routing algorithm} used. When the
total number of arrived and newly created packets to be delivered at
interface $v_i$ exceeds $C(v_i)$, the packets are stored in this
interface's queue and will be processed in the following time steps
on a first-in-first-out(FIFO) basis. If there are several candidate
paths for one packet, one is chosen randomly. Each interface has a
queue for delivering newly arriving packets. Packets reaching their
destinations are deleted from the system. As in
\cite{efficient-routing, edge-deletion,
onset-traffic-congestion,Danila06}, interface buffer size in this
traffic-flow model is set as infinite as it is not relevant to the
\emph{occurrence} of congestion.

For small values of the packet generating rate $R$, the number of
packets on the network is small so that every packet can be
processed and delivered in time. Typically, after a short transient
time, a steady state for the traffic flow is reached where, on
average, the total numbers of packets created and delivered are
equal, resulting in a free-flow state. For larger values of $R$, the
number of packets created is more likely to exceed what the network
can process in time. In this case traffic congestion occurs. As $R$
is increased from zero, we expect to observe two phases: free flow
for small $R$ and a congested phase for large $R$ with a phase
transition from the former to the latter at the critical
packet-generating rate $R_c$.

In order to measure $R_c$, we use the order
parameter~\cite{Arenas01} $ \eta=\lim_{t\rightarrow \infty}{{\langle
\Delta\Theta \rangle} \over {R\Delta t} }$, where $\Theta(t)$ is the
total number of packets in the network at time $t$,
$\Delta\Theta=\Theta(t+\Delta t)-\Theta(t)$,  and
$\langle\cdots\rangle$ indicates the average over time windows of
$\Delta t$. For $R <R_c$ the network is in the free-flow state, then
$\Delta\Theta\approx0$ and $\eta\approx0$; and for $R>R_c$,
$\Delta\Theta$ increases with $\Delta t$ thus $\eta>0$. Therefore in
our simulation we can determine $R_c$ as the transition point where
$\eta$ deviates from zero.

\begin{table}[!t]
\renewcommand{\arraystretch}{1.3}
\centering
 \caption{Elementary topological properties of the five
networks, each containing 1200 nodes. $N$ and $M$ denote the number
of nodes and number of edges respectively, $D$ denotes the diameter,
$L$ denotes the average shortest path length. } \label{basic_info}
\begin{tabular}{ccccc}
\hline Network & $N$ & $M$ & $D$ & $L$
\\
\hline\hline
WS & 1200 & 2400 & 15.5 & 7.86 \\
ER & 1200 & 2450 & 11 & 5.23  \\
BA & 1200 & 2390 & 8 & 4.43  \\
PA & 1200 & 2400 & 8.7 & 4.03  \\
HOT & 1200 & 2583 & 9 & 5.16  \\
  \hline
\end{tabular}

\end{table}

\section{Network designing choices and
objectives}\label{designing-objective-choice}
Of the various network designing choices, the following three ones
are of critical relevance: routing algorithm, link bandwidth
allocation scheme, and network topology. Designers could be faced
with different scenarios characterized by the choices he can make.
For example, when designing a brand new network from the very
beginning, designers own the full freedom of choosing each
ingredient, while when improving an existing network, designers can
only change the routing algorithm or upgrade the link bandwidth.

A specific designing is often associated with explicit designing
objectives. In this paper, we consider three major network designing
objectives: the network transmission capacity, which should be as
large as possible, the maximal bandwidth required to realize the
designing scheme, which should be as small as possible, and the
scalability, which should be as scalable as possible.

Enhancing network transmission capacity is one of the major goals
for network designing and engineering, and is typically measured by
the critical packet-generating rate $R_c$ \cite{efficient-routing,
edge-deletion, onset-traffic-congestion,Danila06}. Although a
simulation-based approach is feasible to measure $R_c$, it is
time-consuming. The betweenness centrality provides a means to
analytically evaluate the $R_c$ under the shortest path routing
algorithm.

In our modified model, for any given network under shortest path
routing, the expected number of packets passing through an edge is
$\frac{RB(e)}{N(N-1)}$ at one time step, with each half flowing
towards one direction, where B(e) is the betweenness centrality of
the edge $e$. So for interface $v_i$ not to get congested, it
follows that $\frac{1}{2}\cdot\frac{RB(e)}{N(N-1)}\leq C(v_i)$,
which leads to $R\leq \frac{2C(v_i)N(N-1)}{B(e)}$. As a result, the
critical packet generating rate $R_c$ for shortest path routing is:
\begin{equation}
\label{rc_theory_equation} R_c=min_{e\in E}\frac{2C(e)N(N-1)}{B(e)}
\end{equation}
where $E$ corresponds to the edge set of the network.

In the more general situation, for any
\emph{topology-based}\footnote{\emph{topology-based} routing
algorithm means routing decision is made solely on the static
topological information, not on dynamic traffic information.}
routing algorithm $\Gamma$, we introduce the effective betweenness
$B^{\Gamma}(e)$ (similar to the definition in
\cite{efficient-routing}) to estimate the possible traffic passing
through an edge under routing algorithm $\Gamma$, which is formally
defined as:
\begin{equation*}
B^{\Gamma}(e)=\sum_{u, v\in V, u\neq
v}\frac{\delta^{\Gamma}_{(e)}(u,v)}{\delta^{\Gamma}(u,v)}
\end{equation*}
where $V$ is the node set, $\delta^{\Gamma}(u,v)$ is the total
number of candidate paths between node $u$ and $v$ under routing
algorithm $\Gamma$, and $\delta^{\Gamma}_{(e)}(u,v)$ is the number
of candidate paths that pass through edge $e$ between $u$ and $v$
under routing algorithm $\Gamma$.

Following this definition, the critical packet generating rate $R_c$
under routing algorithm $\Gamma$ can be calculated as:
\begin{equation}\label{fomular}
R_c=min_{e\in E} \frac{2C(e)N(N-1)}{B^{\Gamma}(e)}
\end{equation}

This equation also explains how different designing ingredients
affect the $R_c$. $C(e)$ represents the bandwidth allocation scheme,
and $B^{\Gamma}(e)$ is the collective effect of network topology and
routing algorithm.

 The maximal bandwidth, denoted as $C_{max}$, required to realize the $R_c$ is another
designing objective of particular importance to network designing
and engineering, because it directly relates to the technical
feasibility(sometimes also monetary issue) of the specific designing
scheme. By posing an upper bound of the required link bandwidth, it
can be used to judge whether the proposed designing scheme can be
realized with state-of-art technologies or available monetary
budget.

Scalability is an issue that becomes increasingly important because
today's networks are all \emph{large-scale} and constantly
\emph{evolving}. A scalable designing will have long-term benefit
for the network investors and operators. In this study, the
scalability will be measured by the growth trends of $R_c$ and
$C_{max}$.

\begin{table*}[!t]
\renewcommand{\arraystretch}{1.3}
\centering
 \caption{Theoretical result of the critical
packet-generating rate $R_c$ under different combinations for
networks in Table \ref{basic_info}. The result is obtained from
Equation \ref{fomular}. For each kind of network, we generate 10
instances, and the result is the average of the 10 instances. }
\label{theoretical_result}
\begin{tabular}{c|ccccc}
\hline

(bandwidth allocation scheme, routing algorithm) & BA & PA & HOT &
ER
&WS\\
\hline \hline

(UC, SPR) & 88.5   & 92.8   & 99.1   & 284.5 & 111.1\\
(UC, EFR) & 264.3  & 195.3  & 94.6   & 390.6 & 147.9\\
(BC, SPR) & 1079.4 & 1192.2 & 1001.7 & 937.9 & 610.2\\
(EBC, EFR)& 766.6  & 844.9  & 909.2  & 891.4 & 606.2 \\
 \hline
\end{tabular}
\end{table*}

\begin{table*}[!t]
\renewcommand{\arraystretch}{1.3}
\centering \caption{Simulation result of the critical
packet-generating rate $R_c$ under different combinations for BA,
PA, HOT, ER and WS networks. For each kind of network, 10 instances
are generated and for each instance, we run ten simulations. The
result here is the average over all the simulations. Compared with
Table \ref{simulation_result}, it is shown that the simulation
result is roughly consistent with the theoretical analysis. }
\label{simulation_result}
\begin{tabular}{c|ccccc}
\hline

(bandwidth allocation scheme, routing algorithm) & BA & PA & HOT & ER &WS\\
\hline \hline

(UC, SPR)  & 110.2 & 115.6  & 119.9  & 370.5 & 136.7 \\
(UC, EFR)  & 320.4 & 249.6  & 114.6  & 519.8 & 174.7 \\
 (BC, SPR) & 934.2 & 1042.7 & 840.8  & 844.2 & 517.0 \\
 (EBC, EFR)& 678.0 & 741.0  & 768.2  & 809.9 & 519.9 \\
 \hline
\end{tabular}
\end{table*}

\begin{table*}[!t]
\centering \caption{$C_{max}$ under different combinations of
network topologies, routing algorithms and bandwidth allocation
schemes. It is evident that betweenness based link bandwidth
allocation scheme often incurs high $C_{max}$, especially for
heterogenous networks. Although (EBC,EFR) can assure much lower
$C_{max}$ for BA-like scale-free networks, the reduction is not
apparent for HOT network. }  \label{max_capability}
\begin{tabular}{c|ccccc}
\hline

(bandwidth allocation scheme, routing algorithm) & BA & PA & HOT &
ER & WS \\
\hline \hline

(UC, *)   & 1     & 1     & 1     & 1     & 1 \\
(BC, *)   & 12.43 & 12.99 & 10.30 & 3.37  & 5.63\\
(EBC, EFR)& 2.91  & 4.45  & 9.70  & 2.40  & 4.13\\

 \hline
\end{tabular}
\end{table*}

Two typical routing algorithms are investigated in this paper: the
shortest path routing, abbreviated as SPR, and the efficient
routing\cite{efficient-routing}, abbreviated as EFR. More formally,
the efficient routing chooses a path $s=v_0,v_1,v_2,\cdots,v_k=t$
between $s$ and $t$ that minimizes the objective function
$\sum_{0\leq i<k}d(v_i)$, where $d(v_i)$ is the vertex degree of
$v_i$.

Three link bandwidth allocation schemes are analyzed: the uniform
link bandwidth allocation scheme, denoted as UC, the betweenness
based link bandwidth allocation scheme, denoted as BC , and the
effective betweenness based link bandwidth allocation scheme,
denoted as EBC. In UC, each link has the same bandwidth, while in BC
and EBC, each link's bandwidth is proportional to its edge
betweenness and effective betweenness respectively. For the purpose
of comparing between different bandwidth allocation schemes, we keep
the condition that the total link bandwidth assigned to all edges
remains fixed, which is set to the number of edges in the network,
i.e., $\sum_eB(e)=M$. The BC scheme is also investigated in
\cite{onset-traffic-congestion}, but it is not treated as an
optimization problem because the total link bandwidth is not fixed.

Thus, if the network topology is predetermined, a designing scheme
is a combination of routing algorithm and link bandwidth allocation
scheme. When network topology is undetermined, it becomes another
ingredient of the designing scheme. We investigate five network
topologies in this paper: BA\cite{BA}, HOT\cite{HOT},
ER\cite{erdos59}, WS\cite{small-world1998}, and PA.  BA network is
constructed according to the standard BA model with $m=2$. ER
network is constructed by the $G_n(p)$ model with the constraint of
connectedness. WS network is built from the ring by randomly
rewiring 15 percent of its edges, also with the constraint of
connectedness. PA is indeed a variant of the BA model, and is
generated by the following process: begin with 3 fully connected
nodes, and add one new node to the graph in successive steps, such
that this new node is connected to the existing nodes with
probability proportional to the current node degree, and finally,
add some internal edges to augment the graph by selecting both
endpoints with probability proportional to the current node degree.
The main difference between PA and BA is that PA has the rich-club
structure\cite{rich-club}, while BA does not.  Finally, the HOT
network is a heuristically optimal topology for the Internet
router-level network, which can be roughly partitioned into three
hierarchies: the low degree core routers, the high degree gateway
routers hanging from the core routers, and the low degree periphery
nodes connected with the gateway routers. The HOT networks generated
here follow the same degree distributions as the corresponding PA
networks.  The basic network properties of these networks of size
1200 are presented in Table \ref{basic_info}.

\section{The cartesian coordinate system and achievable
areas}\label{coordinate-system}
 For a given network, the effect of a
designing scheme can be mapped to a coordinate in the cartesian
coordinate system, with $C_{max}$ being the $X$-axis and $R_c$ being
the $Y$-axis. A natural question is what are the value domains of
$C_{max}$ and $R_c$.

It is easy to see that $C_{max}$'s value range is [1, M]. The
optimal (smallest) value is achieved with uniform bandwidth
allocation scheme, and the largest is achieved by allocating all the
bandwidth to a single edge.

Regarding $R_c$, the lower bound is 0, which can be achieved by some
loop based routing algorithms, that is, the packets are always
looping in the network.

The upper bound $R_c$ is given by the following Theorem.
\newtheorem{thm}{Theorem}
\begin{thm}
\label{theorem} Given a network $G$, the upper bound $R_c$ for any
designing scheme is $2M/L$, where $L$ is the average shortest path
length of $G$; and, this upper bound is only achieved with (BC,
SPR).
\end{thm}
\begin{IEEEproof}
The general idea of the proof is that the number of packets
generated at each time step cannot exceed the maximal number of
packets the network can consume at each time step. Since the total
link bandwidth is $M$, the sum of interface bandwidth is $2M$, which
means the network can at most forward $2M$ packets one step towards
their destinations. Because each packet is generated with random
source and destination, the expected number of steps needed to move
one packet from the source to its destination under routing
algorithm $\Gamma$ is $L^{\Gamma}$, which is equivalent to say that
the network can consume at most $2M/L^{\Gamma}$ packets at each time
step. So, we have
\begin{equation}
\label{inequality} R_c\le 2M/L^{\Gamma} \le 2M/L \end{equation}
which means $2M/L$ is an upper bound of $R_c$.

Then we prove that (BC, SPR) can indeed achieve this upper bound
$R_c$. Recall that $R_c=min_{e\in E}\frac{2C(e)N(N-1)}{B(e)}$, which
in (BC, SPR), can be rewritten as:
\begin{equation*}
\begin{split}
R_c& =min_{e\in E} \frac{2C(e)N(N-1)}{B(e)} \\
&= min_{e\in E} \frac{2B(e)/\sum_{f\in E}B(f)\times M\times N(N-1)}{B(e)} \\
&=\frac{2MN(N-1)}{\sum_{f\in E}B(f)}\\
&=\frac{2MN(N-1)}{N(N-1)L}\\
&=\frac{2M}{L}
\end{split}
\end{equation*}
This proves that the upper bound $2M/L$ is achievable.

Finally, we prove that (BC,SPR) is the only scheme that achieves
$R_c=2M/L$. According to the Inequality \ref{inequality}, only when
two conditions hold simultaneously can $R_c$ reach $2M/L$. These two
conditions are: (1)$L^{\Gamma}=L$, which implies that the routing
algorithm is SPR, and (2) the network can on average move $2M$
packets one step toward their destinations.  Because in SPR, the
expected number of packets arriving at interface $v_i$ in free flow
state is exactly $\frac{1}{2}\frac{RB(e)}{N(N-1)}$, where $e$ is
incident edge of interface $v_i$. When $R=2M/L$, the average total
number of packets in the network is $\sum_e\frac{RB(e)}{N(N-1)}=2M$.
So, for the whole network to move $2M$ packets at each time step,
each interface should exactly move $\frac{MB(e)}{N(N-1)L}$ packets
at each time step, which corresponds to the BC link bandwidth
allocation scheme.
\end{IEEEproof}

In fact, (BC, SPR) reflects the conventional basic philosophy of
real-world empirical network designing. Real-world network routing
protocols such as OSPF and BGP are often based on shortest path
routing(or at least making path length a major decision factor in
path selection). High bandwidth links are placed at key points of
the network, and when some links get overburdened, they will be
upgraded by links with higher bandwidth, potentially attracting even
more traffic and incuring a vicious cycle in traffic demanding and
link bandwidth upgrading.

\begin{figure*}[!t]
\subfigure[BA]{
\includegraphics[width=8cm]{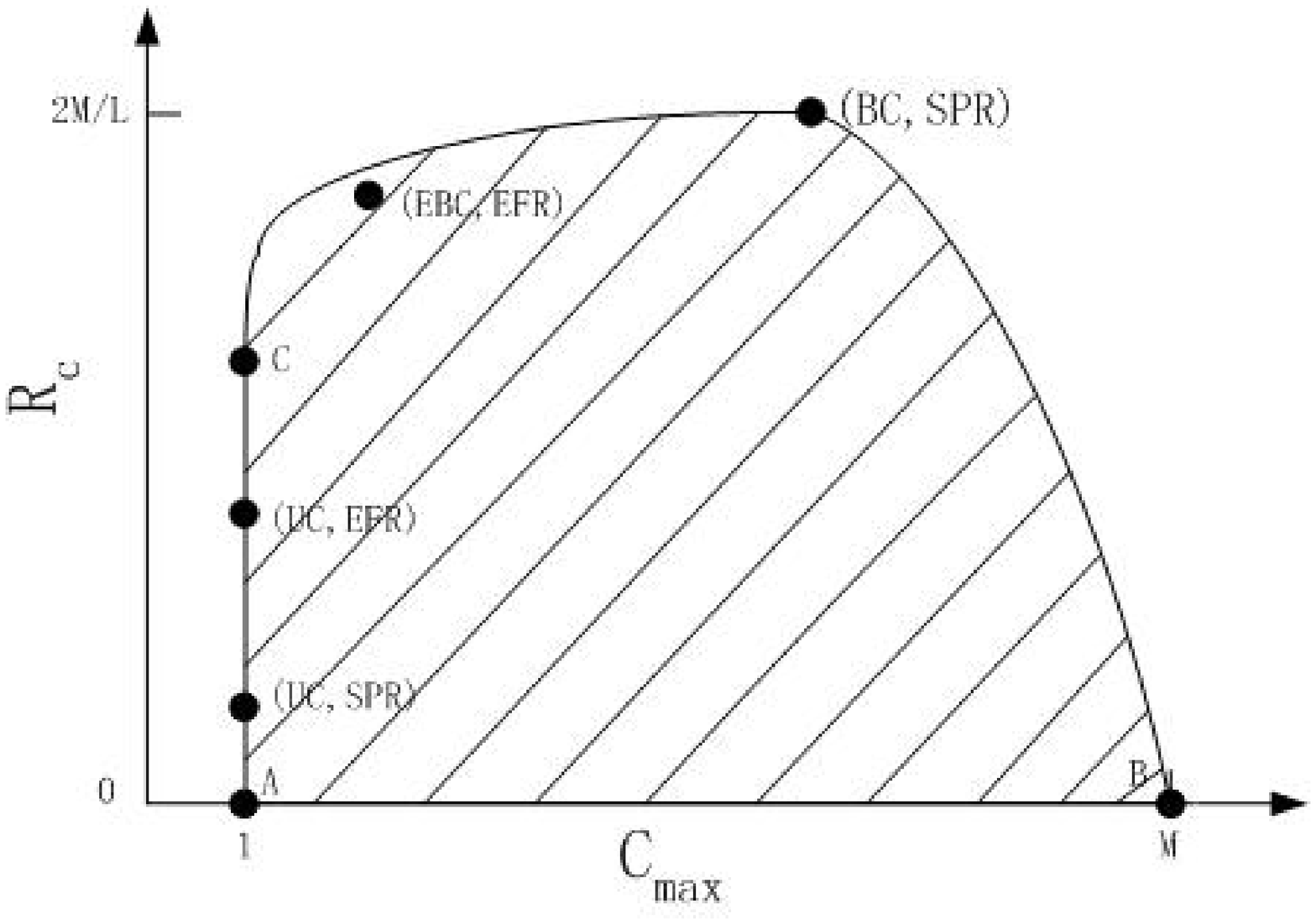}\label{design-space-ba}
} \subfigure[Regular Networks]{
\includegraphics[width=8cm]{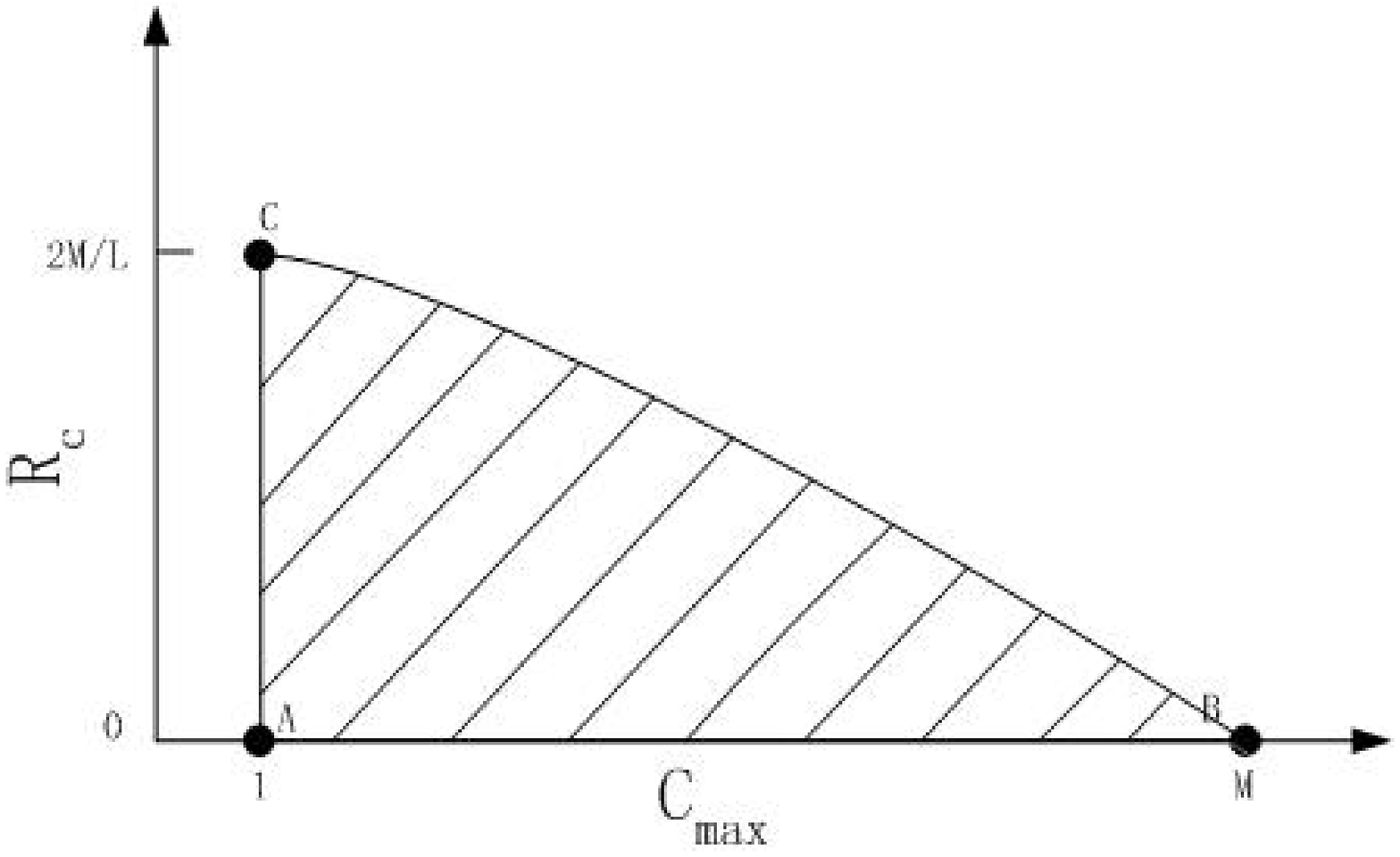}
 \label{design-space-regular}}
\caption{Illustration of the achievable areas for (a) BA-like
networks, and (b) regular networks}
\label{design-space-illustration}
\end{figure*}

\begin{figure*}[!t]
\centering
 \subfigure[$R_c$]{
\includegraphics[width=8cm]{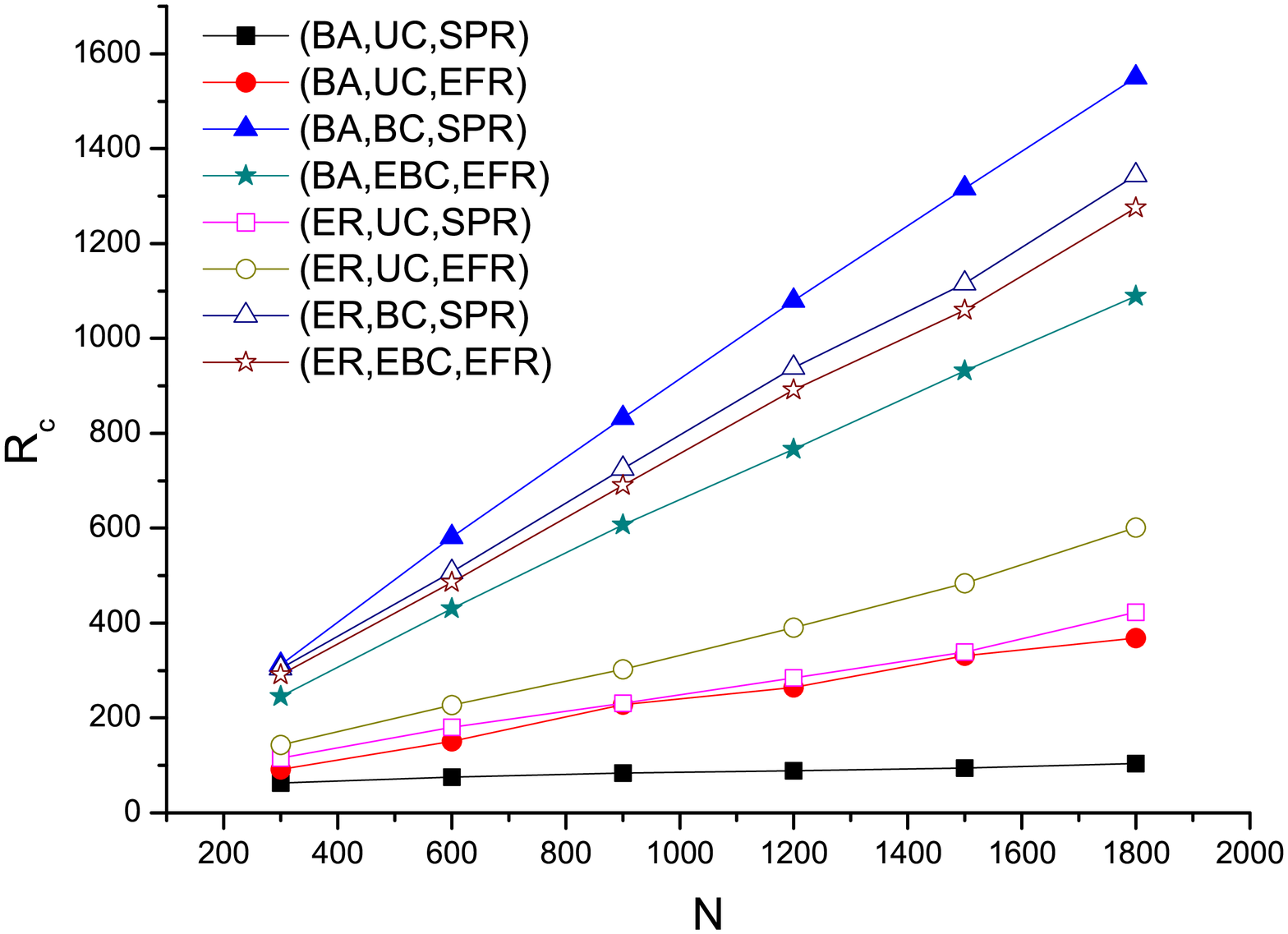}
 \label{rc_evolve}}
\subfigure[$C_{max}$]{
\includegraphics[width=8cm]{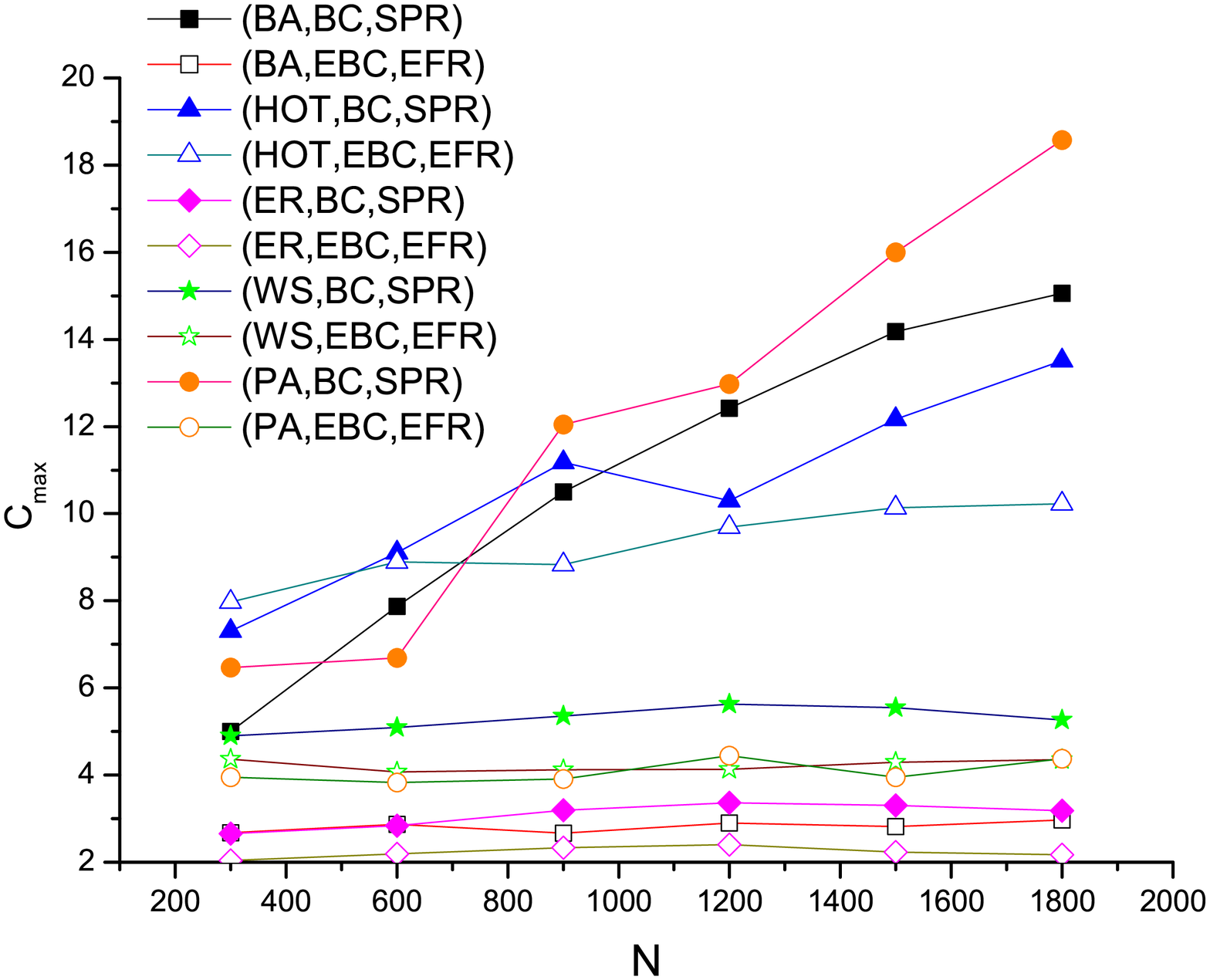}
\label{cmax_evolve}} \caption{(Color online) Scalability of (a)$R_c$
values of BA, HOT and ER under four different configurations; and
(b) $C_{max}$ values of BA and ER under (BC,SPR) and (EBC, EFR)
schemes.} \label{evolvability}

\end{figure*}
Theorem \ref{theorem} also implies that the optimal $R_c$ and
optimal $C_{max}$ cannot be achieved at the same scheme, except for
complete homogenous regular networks, indicating the presence of a
\emph{tradeoff} issue, which is confirmed by Table
\ref{theoretical_result}(with the simulation result of $R_c$
presented in Table \ref{simulation_result}) and Table
\ref{max_capability}.

It can be seen that although (BC,SPR) can guarantee highest $R_c$,
it often incurs high $C_{max}$, because
$C_{max}=\frac{MB_{max}}{N(N-1)L}$ is proportional to the largest
betweenness, which is especially large for heterogenous networks.
Hence, (BC,SPR) is not a cost-effective design for heterogenous
networks. A consequently natural question is what is the achievable
$R_c$ with the lowest $C_{max}$.

With (UC, SPR), it is easy to show that
$R_c=\frac{2N(N-1)}{B_{max}}$. For regular networks such as ring and
toroidal lattice, (UC, SPR) achieves the same $R_c$ as (BC,SPR),
however, for heterogenous networks, $B_{max}$ can be quite large so
that its $R_c$ will be significantly dwarfed compared with $R_c$ in
(BC, SPR)(see Table \ref{theoretical_result} and Fig.
\ref{rc_evolve}).

In BA-like scale-free networks, this $R_c$ can be improved by
changing the routing algorithm from SPR to EFR (see Table
\ref{theoretical_result} and Table \ref{simulation_result}). Indeed,
the $R_c$ in (UC,EFR) is $R_c = \frac{2N(N-1)}{B^{(EFR)}_{max}}$,
where $B^{(EFR)}_{max}$ is the largest effective betweenness
centrality among all edges. While EFR can significantly improve
BA-like network's transmission capacity because $B^{(EFR)}_{max}$ is
substantially decreased compared with $B_{max}$, it is ineffective
for the more realistic HOT model and WS network, and less effective
for the ER network(see Table \ref{theoretical_result}).

It is evident that there is an optimal $R_c$ at the lowest
$C_{max}$. It is easy to demonstrate the existence of such a routing
algorithm that will realize this $R_c$, but it is not an easy task
to find a practically implementable routing algorithm. Recall that
with UC, $R_c=\frac{2N(N-1)}{B^{\Gamma}_{max}}$, so the essence to
realize the optimal $R_c$ at UC is to find a routing algorithm that
minimizes the $B^{\Gamma}_{max}$, which is equivalent to find a path
set of size $N(N-1)$, containing exactly one path for each node
pair, such that the maximal per edge occurrence in this path set is
minimized. The intended path set can be found by firstly computing
all the simple paths between any node pairs, and then enumerating
all combinations of such path set. However, this approach in reality
is infeasible for even median sized networks. A greedy heuristic
algorithm that works can be as follows:

 \begin{enumerate}
 \item
compute all the simple paths between any node pairs;
\item construct
an initial path set, for example, the path set induced by the
shortest path routing or efficient routing;
\item for any node pair, if there is a path such that if replacing this path with the
present one in the path set will reduce the maximal per edge
occurrence, replace the present path with this path. Repeat this
step until there is no such path.
\end{enumerate}

In reality, this routing algorithm is impractical. The significance
of this scheme is that it is a key point in shaping the skeleton of
the achievable area.

An illustration of the achievable area  for BA-like scale-free
networks is shown in Fig. \ref{design-space-ba}, in which the four
typical designing schemes are marked. The other three points A, B
and C, which are important for a conceptual bounding of the
achievable area are also illustrated, where A is a scheme that
combines uniform link bandwidth allocation scheme and some routing
algorithm that keeps the packet always looping around the network, B
is a scheme that allocates all the link bandwidth to a single edge
so that the network's transmission capacity also approaches zero,
and C is a scheme whose existence is mentioned above that realizes
the optimal $R_c$ value with uniform link bandwidth allocation
scheme.

Conceptually, A, B, C and (BC, SPR) together roughly define the
skeleton of the achievable area. However, the shapes may vary
dramatically for different networks. For example, in regular
networks such as the ring or toroidal lattice, the achievable area
for network designing schemes looks like
Fig.\ref{design-space-regular}, in which all the four schemes
discussed in this paper as well as C will converge to the single
point C. And (UC,EFR) will not always top (UC,SPR), such as in WS
network.

From an engineering perspective, a relevant problem is whether there
exists a designing scheme that achieves good tradeoff between $R_c$
and $C_{max}$. The answer is that this property heavily depends on
the particular network topology. However, in many cases, such
designing exists.  This paper proposes one such designing, (EBC,
EFR), which proves to be cost-effective for most small-world
networks, especially for BA-like scale-free networks.

To see why this scheme has the potential of achieving good tradeoff,
first note that with (EBC, EFR), $R_c=\frac{2M}{L^{(EFR)}}$ and,
$C_{max}=\frac{MB^{(EFR)}_{max}}{N(N-1)L^{(EFR)}}$. So, we have:

\begin{equation*}
\frac{R^{(BC,SPR)}_c}{R^{(EBC,EFR)}_c}=\frac{L^{(EFR)}}{L}
\end{equation*}

and
\begin{equation*}
\frac{C^{(BC,SPR)}_{max}}{C^{(EBC,EFR)}_{max}}=\frac{L^{(EFR)}}{L}\times\frac{B_{max}}{B^{(EFR)}_{max}}
\end{equation*}

\begin{figure*}[!t]
\centering
 \subfigure[$B_{max}$]{
\includegraphics[width=8cm]{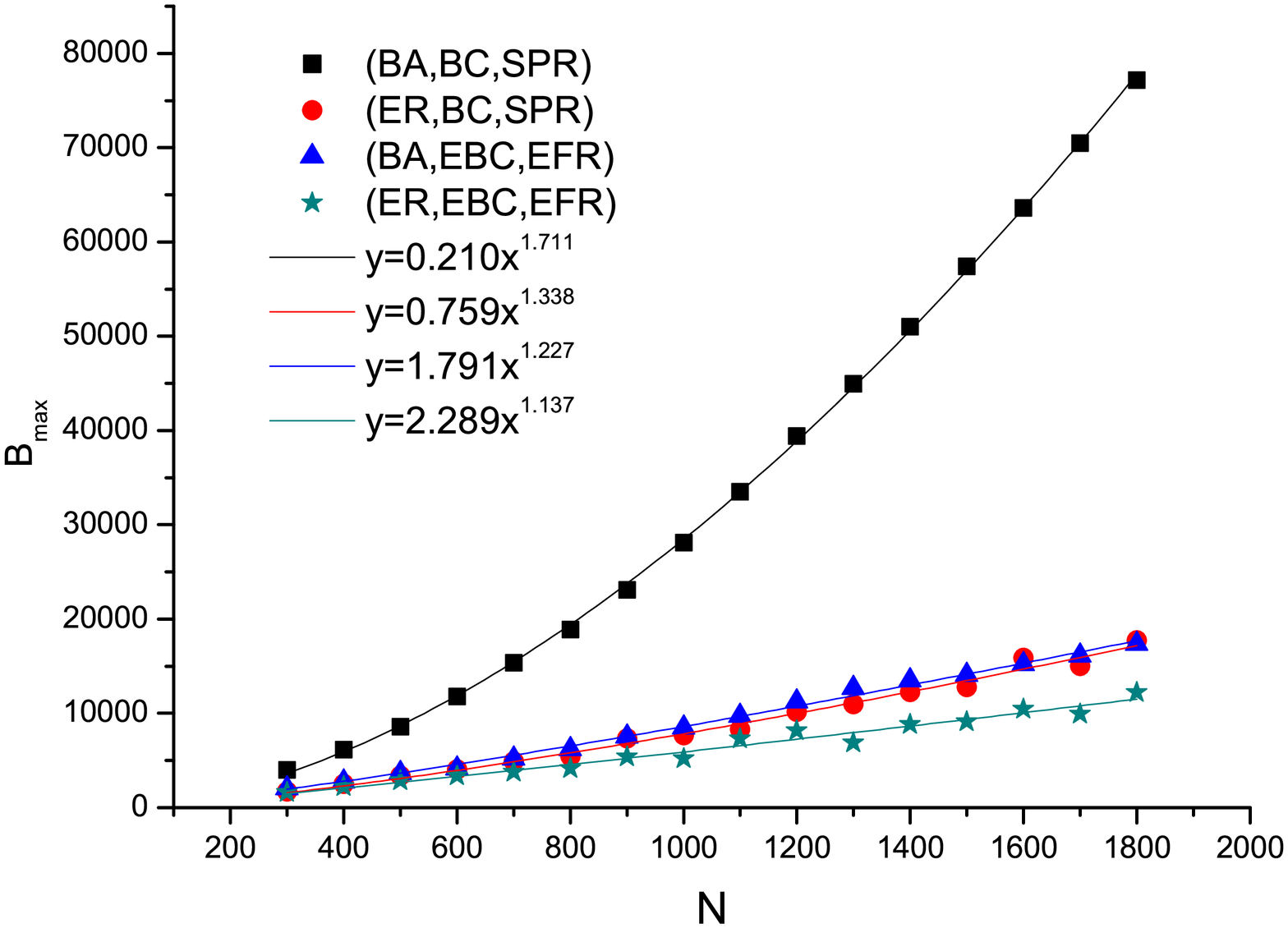}
\label{bmax_scale}} \subfigure[APL]{
\includegraphics[width=8cm]{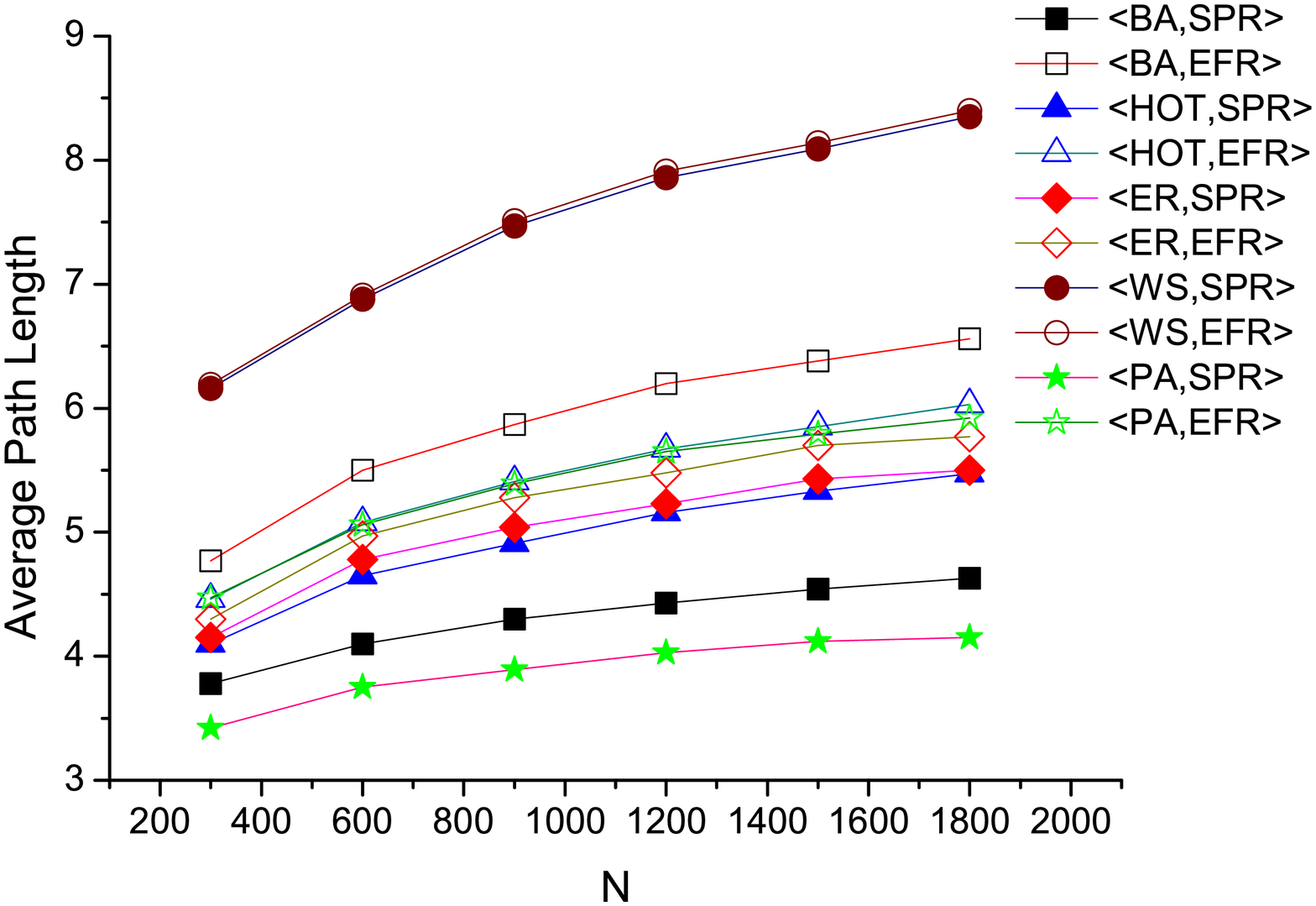}
 \label{apl} }
 \caption{(Color online) (a) Scalability of the maximal betweenness and
effective betweenness for BA and ER networks. With SPR, the $Y$-axis
denotes the value of maximal betweenness, and with EFR, the $Y$-axis
denotes the maximal effective betweenness. The line fittings for
these data are also presented, and (b) average path length for the
five networks with different size under both shortest path routing
and efficient routing. The filled shape represents the SPR, while
the hollow shape represents the EFR.}
\end{figure*}
In most networks, $L^{(EFR)}$ is only slightly longer than $L$
\footnote{This also ensures that the path length overhead incurred
is also acceptable for all the networks investigated in this
paper.}(see Fig.\ref{apl} for the average path length of different
routing algorithms in different networks), so the $R_c$ in (EBC,EFR)
is only slightly smaller than the optimal one.   However, $C_{max}$
in (BC,SPR) is
$\frac{L^{(EFR)}}{L}\times\frac{B_{max}}{B^{(EFR)}_{max}}$ times
larger than the $C_{max}$ in (EBC,EFR). This expresses why the
cost-effectiveness of (EBC,EFR) is topology-dependent.

In HOT network, although with the same skewed degree distribution,
its rigid hierarchy structure restricts the flexibility in path
selection, making it insensitive to routing algorithm changes, so
$B_{max}$ and $B^{(EFR)}_{max}$ show no big difference. In WS
network, the near uniform degree distribution makes SPR and EFR
nearly identical routing algorithms, so $B_{max}$ and
$B_{max}^{(EFR)}$ also show no difference. The difference between
HOT and WS is that in HOT network, $C_{max}$ is high, while, in WS,
$C_{max}$ is relatively low.

Scale-free networks such as BA and PA, on the other hand, have the
inherent flexibility in path selection and show strong sensitivity
to routing algorithm changes. In these networks, $B_{max}$ can be an
order of magnitude larger than $B^{(EFR)}_{max}$, which makes
$C_{max}$ in (EBC,EFR) quite smaller compared with (BC,SPR).

ER network's random structure makes it lies somewhere between the WS
and BA network. For ER network, both (BC,SPR) and (EBC,EFR) are good
designing choices. Even (UC,SPR) and (UC,EFR) also achieves
relatively high $R_c$ values compared with other networks.

\section{Scalability of network designing
schemes}\label{scalability} The scalability is investigated by the
growth trends of $R_c$ and $C_{max}$ when network evolves.
Conceptually, the scalability can be viewed as a third dimension of
the designing objective, i.e., the time dimension. Figure
\ref{rc_evolve} presents the $R_c$ values for BA and ER networks
under four different settings and Figure \ref{cmax_evolve} presents
the $C_{max}$ values for the five networks under (BC, SPR) and (EBC,
EFR). It is shown that under (UC, SPR), BA network's $R_c$ value
remains quite stable, i.e., almost not scaling with $N$. This is
because, $B_{max}$ scales super linearly with $N$, as illustrated in
Fig. \ref{bmax_scale}. Numerical fitting on simulation result shows
that $B_{max}\sim N^{1.711}$ in BA network, whereas in ER network,
$B_{max}\sim N^{1.338}$. On the other hand, $B^{(EFR)}_{max}$ grows
much slowly as network expands, $\sim N^{1.227}$ and $\sim
N^{1.137}$ for BA and ER respectively. This means with (UC, SPR), BA
network's $R_c$ scales very slowly, while ER network's $R_c$ scales
much better. However, with (UC, EFR) the scalability of $R_c$
significantly improves for BA network.

Regarding $C_{max}$, with (BC, SPR), $C_{max}\propto B_{max}$, so it
grows fast for BA, HOT and PA networks, and much more slowly for ER
and WS networks. With (EBC, EFR), $C_{max}\propto B^{(EFR)}_{max}$,
which scales much slowly in BA and PA networks, but still grows fast
in HOT network. This is evidenced in Fig.\ref{cmax_evolve}, where
$C_{max}$ grows fast with (BA, BC, SPR), (PA, BC, SPR) and (HOT, BC,
SPR) and (HOT, EBC, EFR), but remains nearly stable for other
settings.

In short, the above analysis show that (BC,SPR) is not a scalable
designing for heterogenous networks. On the other hand, our proposed
(EBC, EFR) is a cost-effective designing that is scalable for most
of the networks with small-world properties, except for heterogenous
networks with rigid hierarchy structure, and ER network shows the
potential of achieving scalable cost-effective designing under any
settings.

 \section{Conclusions}\label{conclusion}
By proposing that network designing is not a single objective
optimization process, but rather a multi-objective optimization
process, this paper established a scientific foundation for network
designing studies. It has been found that typical network designing
objectives such as network transmission capacity and technical
feasibility are often contradictive, which indicates the presence of
a tradeoff issue. We defined a two dimensional cartesian system to
evaluate the effect of each designing scheme.

In particular, the empirical network designing approach widely
adopted today is incapable to meet the cost-effective and scalable
requirements. The existence of cost-effective design depends heavily
on the network topology. For BA-like scale-free networks, efficient
routing combined with effective betweenness based link bandwidth
allocation scheme can be regarded as a cost-effective design, and
this property also scales as network evolves. If the designer has
the full freedom of choosing all designing ingredients, then ER
network shows pretty good characteristic in achieving a
cost-effective design under most settings.

We believe the scientific framework as well as the specific findings
will give insightful help to advance the science of network
designing and engineering. Evaluating the performance of other
designing schemes under this framework, and finding the exact
boundary of the achievable area, for example, whether the boundary
is concave or convex for different networks, are the direction for
future work.
\section*{Acknowledgment}
This work is partly supported by the National Natural Science
Foundation of China under Grant No. 60673168 and the Hi-Tech
Research and Development Program of China under Grant No.
2008AA01Z203.


\begin{thebibliography}{5}

\bibitem{routing-scaling} C. Wittbrodi, B. Woodcock, A. Ahuja, T.
Li, V. Gill and E. Chen, "Global routing system scaling issues",
NANOG Panel, http://www.nanog.org/mtg-0102/witt.html (2001).

\bibitem{huston-1} G. Huston, "Analyzing the Internet's BGP routing
table", The Internet Protocol Journal, {\bf 4} (2001).

\bibitem{huston-2} G. Huston "Scaling inter-domain routing-a view
forward", The Internet Protocol Journal, {\bf 4} (2001).

\bibitem{claffy2003} A. Broido, E. Nemeth and K. Claffy, "Internet
expansion, refinement, and churn", European Transactions on
Telecommunications, {\bf 13}, 33-51 (2002).


\bibitem{powerlaw1999} M. Faloutsos, P. Faloutsos, and C. Faloutsos,
"On power-law relationships of the Internet topology", \emph{Proc.
ACM SIGCOMM 1999} (1999).

\bibitem{BA} A. L. Barab$\acute{a}$si and R. Albert, "Emergence of scaling in random networks",Science {\bf 286}, 509
(1999).

\bibitem{GLP} T. Bu and D. Towsley, "On distinguishing between
Internet power-law topology generators", \emph{Proc. IEEE INFOCOM
2002} (2002).

\bibitem{chinese-internet}
S.~Zhou, G. Q. Zhang, and G. Q. Zhang, ``{Chinese Internet AS-Level
  Topology},'' \emph{IET Commun.}, {\bf 1}, 209--214 (2007).

\bibitem{dk-series} P. Mahadevan, D. Krioukov, K. Fall and A.
Vahdat, "Systematic topology analysis and generation using degree
correlations", \emph{Proc. ACM SIGCOMM 2006} (2006).

\bibitem{internet-evolution}G. Q. Zhang, G. Q. Zhang, Q. F. Yang,
S. Q. Cheng, and T. Zhou, "Evolution of the Internet and its cores",
New. J. Phys. {\bf 10}, 123027 (2008).



\bibitem{netse}Network Seicen and Engineering(NetSE) Research
Agenda, http://www.cra.org/ccc/docs/NetSE-Research-Agenda.pdf

\bibitem{gupte-1} N. Gupte and B. K. Singh,  "Role of connectivity in congestion and decongestion in networks", Euro. Phys. B. {\bf 50}, 227 (2006).
\bibitem{gupte-2}N. Gupte, B. K. Singh and T. M. Janaki, "Networks: structure, function and optimization", Phys. A.
{\bf 346}, 75 (2005).

\bibitem{node-capability-redistribution} G. Q. Zhang, S. Zhou, G.
Yan, D. Wang, and G. Q. Zhang, "Enhancing the network transmission
capability by efficiently allocating node capability",
arXiv:0910.2285 (2009).

\bibitem{cost-effective-designing} G. Q. Zhang, "On cost-effective
communication network designing", arxiv:0910.2104 (2009).

\bibitem{centrality-and-network-flow} S. P. Borgatti,  "Centrality and Network flow", Soc. Netw. {\bf 27}, 55
(2005).

\bibitem{load-distribution} K. -I. Goh, B. Kahng, and D. Kim, "Universal behavior of load distribution in scale-free newtorks", Phys. Rev. Lett. {\bf
87}, 278701 (2001).

\bibitem{optimal-network-topology} R. Guimer$\grave{a}$, A. Z. Guilera, F. V. Redondo, A. Cabrales, and A. Arenas, "optimal network topologies for local search with congestion", Phys. Rev. Lett. {\bf 89}, 328170
(2002).

\bibitem{polymorphic-torus} H. Li and M. Maresca, "Polymorphic-torus network", IEEE. Trans. Computers {\bf 38}, 1345
(1989).

\bibitem{edge-deletion} G. Q. Zhang, D. Wang, and G. J. Li, "Enhancing the transmission efficiency by edge deletion in scale-free networks", Phys. Rev. E {\bf 71}, 017101
(2007).

\bibitem{efficient-routing} G. Yan, T. Zhou, B. Hu, Z. Q. Fu, and B. H. Wang, "Efficient routing on complex networks", Phys. Rev. E {\bf 73}, 046108
(2006).

\bibitem{onset-traffic-congestion} L. Zhao, Y. C. Lai, K. Park, and N. Ye, "Onset of traffic congestion in complex networks", Phys. Rev. E {\bf 71}, 026125
(2005).
\bibitem{assortative-mixing} M. E. J. Newman, "Assortative mixing in networks", Phys. Rev. Lett. {\bf
89}, 208701 (2002).
\bibitem{clustering} S. N. Dorogovtsev, "Clustering of correlated networks", Phys. Rev. E. {\bf 69},
027014 (2004).

\bibitem{ark}Archipelago Measurement Infrasture,
http://www.caida.org /projects/ark/

\bibitem{DIMES} Y. Shavtt and E. Shir, "DIMES-letting the Internet
measure itself", http://www.arxiv.org/abs/cs.NI/0506099 (2005).

\bibitem{rocketfuel} N. Spring, R. Mahajan, D. Wetherall and T.
Anderson, "Measuring ISP topologies with rocketfuel", \emph{Proc.
IEEE INFOCOM 2001} (2001).

\bibitem{Danila06}
B. Danila, Y. Yu, J. A. Marsh, and K. E. Bassler, "Optimal transport
on complex networks",  Phys. Rev. E {\bf 74}, 046106 (2006).


\bibitem{HOT} L. Li, D. Alderson, W. Willinger, J. Doyle, "A first principles approach to understanding the Internet's rouer-level topology", \emph{Proc. ACM SIGCOMM 2004} (2004).

\bibitem{rich-club} S. Zhou and R. J. Mondrag\'on, "The rich-club
phenomenon in the Internet topology", IEEE Commun. Lett. {\emph 8},
180 (2004).
\bibitem{Arenas01}A. Arenas, A. D\'iaz-Guilera, and R. Guimer\`a, "Communication in networks
  with hierarchical branching", Phys. Rev. Lett.
{\bf 86}, 3196 (2001).

\bibitem{freeman-betweenness} L. C. Freeman, "Centrality in social networks: conceptual clarification", Soc. Netw. {\bf 1}, 215 (1979).

\bibitem{erdos59}P. Erd\"{o}s and A. R\'enyi, "On random graphs", Publ. Math. Debrecen
  {\bf 6}, 290 (1959).


\bibitem{small-world1998} D. J. Watts and S. H. Strogatz, "Collective Dynamics of small world networks", Nature {\bf 393},
440 (1998).



\end{thebibliography}
 \end{document}